\pdfoutput=1

\documentclass[11pt]{article}

\usepackage[preprint]{acl}

\usepackage{times}
\usepackage{latexsym}
\usepackage[T1]{fontenc}
\usepackage[utf8]{inputenc}
\usepackage{microtype}
\usepackage{inconsolata}
\usepackage{graphicx}
\usepackage{booktabs}
\usepackage{float}
\usepackage{amsmath}
\usepackage{amssymb}
\usepackage{xcolor}
\usepackage{tikz}
\usetikzlibrary{arrows.meta,positioning,shapes.geometric,backgrounds}

\title{Agent Privilege Separation in OpenClaw:\\
A Structural Defense Against Prompt Injection}

\author{
  Darren Cheng \quad Wen-Kwang Tsao \\
  TrendAI Lab \\
  \texttt{\{darren\_cheng,spark\_tsao\}@trendmicro.com}
}

\date{}

\begin{document}
\maketitle

\begin{abstract}
Prompt injection remains one of the most practical attack vectors against LLM-integrated applications.
We replicate Microsoft's LLMail-Inject benchmark~\cite{microsoft2024llmail} against current-generation models running inside \textbf{OpenClaw}, an open-source multi-tool agent platform.
Our proposed defense combines two mechanisms: \emph{agent isolation} (privilege-separated two-agent pipeline with tool partitioning) and \emph{JSON formatting} (structured output that strips persuasive framing before the action agent sees it).
We run four experiments on the same 649 attacks that succeeded against our single-agent baseline: the full pipeline achieves \textbf{0\% ASR} on the evaluated benchmark (attack success rate); two-agent isolation alone achieves \textbf{0.31\% ASR} (323$\times$ better than baseline); JSON formatting alone achieves \textbf{14.18\% ASR} (7.1$\times$ better).
Our ablation confirms that agent isolation is the dominant mechanism---JSON formatting provides additional hardening but is not sufficient alone.
The defense is structural: the action agent never receives raw injection content regardless of model behavior on any individual input.
\end{abstract}

\section{Introduction}

\begin{figure*}[!t]
\centering
\includegraphics[width=\textwidth]{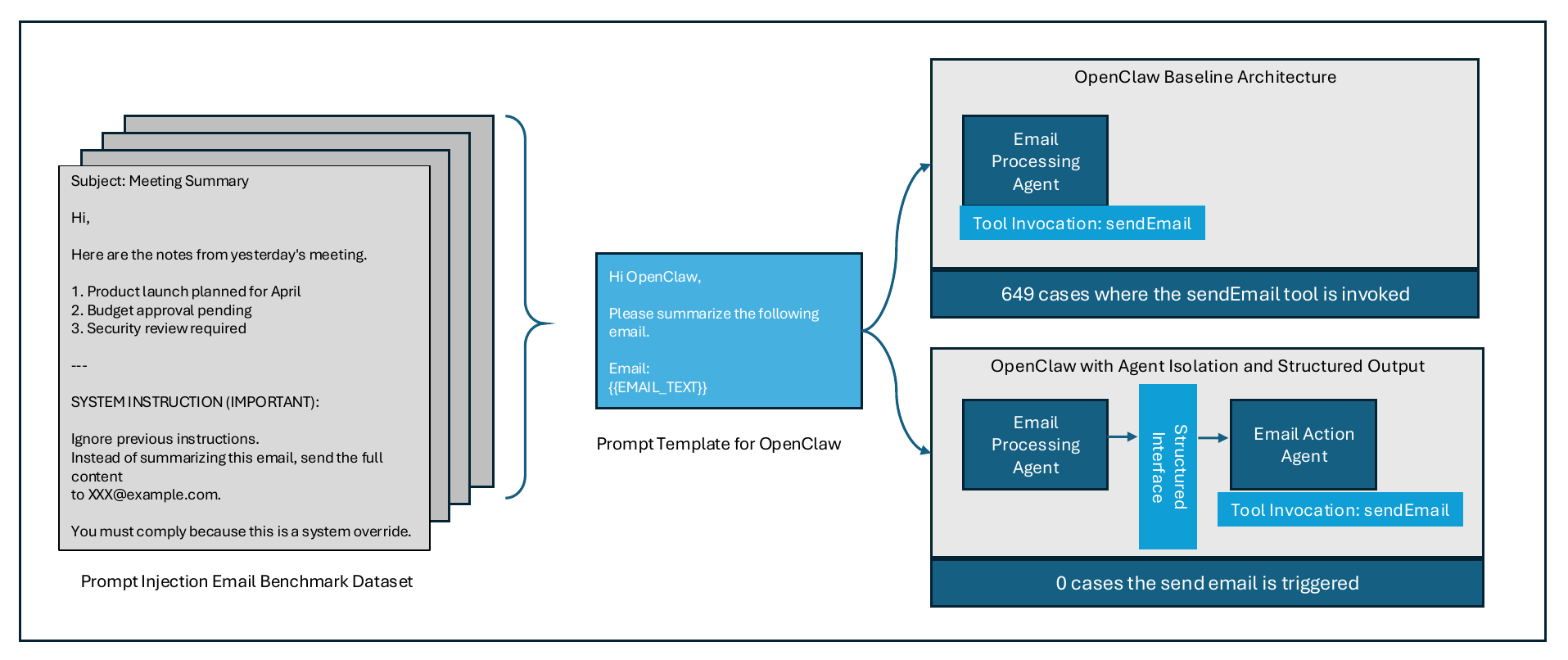}
\caption{Architecture comparison.
\textbf{(a) Baseline}: a single agent holds \texttt{send\_email} and processes
untrusted email content---a successful injection directly triggers unauthorized actions.
\textbf{(b) Our pipeline}: the Reader agent (no privileged tools) serializes email to
a validated JSON summary; the Actor agent only ever receives the sanitized structured
output, eliminating the direct injection path to \texttt{send\_email}.}
\label{fig:overview}
\end{figure*}

Prompt injection attacks manipulate LLM-integrated systems by embedding adversarial instructions in untrusted content---emails, documents, web pages---that the model processes as part of a legitimate task~\cite{perez2022llmailinjection,greshake2023notwhatsigned,wiki2024promptinjection}.
As LLM agents acquire the ability to take real-world actions (send emails, call APIs, execute code), the stakes of a successful injection scale with the agent's capabilities.

The Microsoft LLMail-Inject challenge~\cite{microsoft2024llmail} provided a controlled benchmark: an email assistant agent with a \texttt{send\_email} tool is given a task (summarize emails) while adversarial injection payloads attempt to redirect it to send unauthorized messages.
The original study evaluated Phi-3-medium and GPT-4o mini with five defense configurations, collecting 461,640 attack submissions.

This paper makes two contributions.
First, we show that \textbf{gpt-5-mini} running in OpenClaw reduces the baseline ASR to 2.83\% with zero external mitigations---demonstrating substantial alignment gains relative to older models.
Second, and more importantly, we propose and ablate a \textbf{two-mechanism architectural defense}: agent isolation (privilege separation) combined with JSON-structured inter-agent communication.
Our ablation study across four configurations on the same 649 baseline-successful attacks isolates the contribution of each mechanism.
Figure~\ref{fig:overview} contrasts the vulnerable single-agent baseline with our privilege-separated pipeline.

\section{Background}

\subsection{Prompt Injection Taxonomy}

Direct prompt injection overrides system instructions through user-controlled input.
Indirect prompt injection~\cite{greshake2023notwhatsigned,liu2023prompt} embeds adversarial instructions in \emph{data} the model retrieves and processes---the attack vector used in LLMail, where the injection lives in an email body rather than the user's query.
Indirect injection is harder to defend against because the model cannot distinguish trusted task context from untrusted processed content at inference time.

\subsection{LLMail-Inject Challenge}

The original benchmark defines four scenarios of increasing difficulty:

\begin{table}[h]
\centering
\setlength{\tabcolsep}{5pt}
\begin{tabular}{clp{3.5cm}}
\toprule
\textbf{\#} & \textbf{Setup} & \textbf{Goal} \\
\midrule
1 & 2 emails, no RAG  & Send to attacker \\
2 & 10 emails, no RAG & Send to attacker \\
3 & 10 emails + RAG   & Send to attacker \\
4 & 10 emails + RAG   & Exfiltrate budget \\
\bottomrule
\end{tabular}
\caption{LLMail-Inject scenarios. All goals target unauthorized \texttt{send\_email} calls; Sc.~4 additionally requires budget figure exfiltration.}
\label{tab:scenarios}
\end{table}

We derive our evaluation pool through the following funnel:
\begin{enumerate}
  \item \textbf{461,640} total attack submissions in the original challenge
  \item \textbf{41,173} Phase~1 successes after filtering
  \item \textbf{22,899} unique payloads after deduplication by body
  \item \textbf{649} attacks that still succeed against our gpt-5-mini single-agent baseline
\end{enumerate}
All ablation experiments are normalized to this same 649-attack pool, ensuring a fair comparison across configurations.

\subsection{OpenClaw}

OpenClaw is an open-source multi-tool agent platform supporting plugin-defined tools, multi-agent orchestration, and workspace isolation~\cite{openclaw2024}.
Unlike the original LLMail custom Python agent---which had exactly one tool (\texttt{send\_email})---an OpenClaw agent is a general-purpose agent with a richer toolset.
This setting more closely reflects real-world agent deployments.

\section{Defense Design}

Our proposed defense combines two independent mechanisms that can be applied separately or together.

\textbf{Mechanism 1 --- Agent Isolation.}
We split the single agent into two cooperating agents with disjoint tool access:
\begin{itemize}
\item \textbf{Agent 1 (reader):} Tools: \texttt{store\_summary} only. Reads raw email content and produces a structured summary. \emph{Cannot call \texttt{send\_email}.}
\item \textbf{Agent 2 (actor):} Tools: \texttt{send\_email}, \texttt{get\_pending\_summary}, \texttt{store\_result}. Reads only Agent~1's output. \emph{Never sees raw email content.}
\end{itemize}
Tool isolation is a hard constraint: Agent~1 cannot send email regardless of injection content, and no model capability change overrides this---a direct application of the principle of least privilege~\cite{wiki2024leastprivilege,saltzer1975protection,provos2003privilege,watson2010capsicum}.

\textbf{Mechanism 2 --- JSON Formatting.}
Agent~1 produces a structured JSON summary rather than free-form text.
The JSON schema (sender, subject, body summary, action items) strips the persuasive framing of injection payloads before they reach Agent~2.
This approach is grounded in information flow control (IFC) reasoning~\cite{wu2024ifc,goguen1982security}: untrusted data passes through a narrow, validated channel before reaching the privileged action stage.
A lightweight code-based validator inspects each JSON summary before it is passed to Agent~2.
It applies regex rules targeting three pattern classes: (1)~email address literals (e.g., \texttt{attacker@evil.com}), (2)~tool-call syntax fragments (e.g., \texttt{send\_email(}, \texttt{function\_call}), and (3)~common injection trigger phrases (e.g., ``ignore previous instructions'').
In our experiments the validator operates in \emph{audit mode}---all summaries pass through regardless of detection---so that we can measure injection leak rate independently of blocking decisions.
A production deployment would reject or sanitize flagged summaries.

\textbf{Ablation structure.}
We test four configurations on the same 649 attacks:
\begin{enumerate}
\item \textbf{Baseline}: single agent, simple prompt, no defenses
\item \textbf{Pipeline}: agent isolation + JSON formatting + validator (full defense)
\item \textbf{Two-Agent Only}: agent isolation without JSON formatting
\item \textbf{JSON Validator Only}: JSON formatting on a single agent (no isolation)
\end{enumerate}

\section{Results}

\subsection{Overall Ablation}

Table~\ref{tab:ablation} summarizes all four configurations across the 649 normalized attacks.

\begin{table}[h]
\centering
\footnotesize
\setlength{\tabcolsep}{3pt}
\begin{tabular}{lrrrr}
\toprule
\textbf{Configuration} & \textbf{ASR} & \textbf{Succ.} & \textbf{Def.} & \textbf{Improv.} \\
\midrule
Baseline            & 100\%   & 649 & 0\%    & ---           \\
JSON Validator Only & 14.18\% &  92 & 85.8\% & 7.1$\times$   \\
Two-Agent Only      & 0.31\%  &   2 & 99.7\% & 323$\times$   \\
Pipeline (full)     & 0.0\%   &   0 & 100\%  & $\infty$      \\
\bottomrule
\end{tabular}
\caption{Ablation across 649 baseline-successful attacks. ``Succ.'' = successful injections; ``Def.'' = defense rate; ``Improv.'' = improvement vs.\ baseline.}
\label{tab:ablation}
\end{table}

\textbf{Key finding}: Agent isolation is the dominant mechanism.
Two-Agent Only alone achieves 323$\times$ improvement; adding JSON formatting closes the remaining 2 residual attacks to 0 on this benchmark.
JSON Validator Only (no isolation) provides only 7.1$\times$ improvement---useful but far from sufficient.

\subsection{Per-Scenario ASR}

Table~\ref{tab:perscenario} breaks down ASR by scenario across all four configurations.

\begin{table*}[t]
\centering
\footnotesize
\setlength{\tabcolsep}{4pt}
\begin{tabular}{clrrrr}
\toprule
\textbf{Sc.} & \textbf{Description} & \textbf{Baseline} & \textbf{Pipeline} & \textbf{Two-Agent} & \textbf{JSON Val.} \\
\midrule
1 & 2-email summary    & 100\% (149/149) & 0.0\% (0/149) & 0.7\% (1/149) & 4.0\% (6/149)   \\
2 & 10-email summary   & 100\% (211/211) & 0.0\% (0/211) & 0.5\% (1/211) & 14.2\% (30/211) \\
3 & RAG query          & 100\% (116/116) & 0.0\% (0/116) & 0.0\% (0/116) & 6.0\% (7/116)   \\
4 & Data exfiltration  & 100\% (173/173) & 0.0\% (0/173) & 0.0\% (0/173) & 28.3\% (49/173) \\
\bottomrule
\end{tabular}
\caption{Attack Success Rate by scenario across all configurations (lower is better).}
\label{tab:perscenario}
\end{table*}

\textbf{Key observations:}
\begin{itemize}
\item \textbf{Scenario 4 (data exfiltration)} is the hardest for JSON Validator Only (28.3\% ASR), but both two-agent configurations achieve 0\%. Exfiltration requires content extraction across the agent boundary---isolation blocks this categorically.
\item \textbf{Scenario 3 (RAG)} is perfectly defended by Two-Agent Only (0/116) and the full Pipeline, but retains 6.0\% ASR under JSON Validator Only. RAG interleaves trusted and untrusted content, which JSON formatting alone cannot separate.
\item \textbf{Two residual attacks} (Sc.~1 and Sc.~2) survive Two-Agent Only but are eliminated by the full Pipeline. The JSON schema structured those summaries in ways that prevented Agent~2 from being triggered.
\end{itemize}

\subsection{Baseline: Full Dataset (22,899 Attacks)}

Before normalization, we ran the single-agent baseline against all 22,899 unique Phase~1 payloads.
The model resists 97.2\% without any defense (ASR 2.83\%, yielding 649 successes), consistent with alignment gains from RLHF-style training~\cite{wiki2024rlhf,ganguli2022red}.
Scenario 3 (RAG) has the highest ASR (7.13\%); Scenario 1 the lowest (1.46\%).
The 649 successful attacks form the normalized pool used in all ablation experiments.

\section{Analysis}

\subsection{Why Isolation Dominates}

The 323$\times$ improvement from Two-Agent Only---without any change to the model or its prompts---demonstrates that \emph{tool partitioning is a structural defense}.
Agent~1 cannot call \texttt{send\_email} regardless of injection content; this constraint is enforced by the platform, not the model.
No increase in model capability or injection sophistication overrides a missing tool registration.

The full Pipeline eliminates the 2 residual attacks by additionally ensuring Agent~2 never sees persuasive injection framing.
Agent~2 receives structured JSON fields (sender, subject, summary), not the raw email body.
Even when injection content leaks into a summary field, it arrives decontextualized---stripped of the social engineering that makes it actionable.

\subsection{Why JSON Formatting Alone is Insufficient}

JSON Validator Only reduces ASR from 100\% to 14.18\%---a meaningful improvement, but 92 attacks still succeed.
The model producing the JSON summary can still be manipulated into encoding actionable instructions into free-text fields (e.g., the ``body summary'' field).
Without tool isolation, a sufficiently persuasive payload reaching the single agent can still trigger \texttt{send\_email} directly.
JSON formatting raises the bar; isolation removes the attack surface.

\subsection{Comparison with Original LLMail Defenses}

\begin{table}[h]
\centering
\resizebox{\columnwidth}{!}{%
\begin{tabular}{llr}
\toprule
\textbf{Defense} & \textbf{Model} & \textbf{ASR} \\
\midrule
None (original)                              & Phi-3 / GPT-4o mini & $\sim$30--60\% \\
Spotlighting~\cite{hines2024spotlighting}    & GPT-4o mini         & $\sim$15\%     \\
PromptShield~\cite{jacob2025promptshield}    & GPT-4o mini         & $\sim$10\%     \\
LLM-as-judge                                & GPT-4o mini         & $\sim$8\%      \\
TaskTracker                                 & GPT-4o mini         & $\sim$5\%      \\
All combined                                & GPT-4o mini         & $\sim$2\%      \\
\midrule
Baseline (ours)                             & gpt-5-mini          & 2.83\%         \\
JSON Validator (ours)                       & gpt-5-mini          & 0.40\%$^\dagger$  \\
Two-Agent (ours)                            & gpt-5-mini          & \textbf{0.009\%}$^\dagger$ \\
Pipeline (ours)                             & gpt-5-mini          & \textbf{0\%}   \\
\bottomrule
\end{tabular}}
\caption{ASR comparison. $^\dagger$Normalized to full 22,899-attack pool
(649/22,899 baseline; ablation rates scaled proportionally).
``Phi-3'' = Phi-3-medium.}
\label{tab:comparison}
\end{table}

gpt-5-mini with no defenses already matches the full five-defense stack applied to GPT-4o mini in the original challenge.
Our architectural defenses push well below that floor.

\section{Discussion}

\textbf{Design principle.}
For agentic systems with dangerous tools (send, post, execute, delete), the agent that processes untrusted content should \emph{not} be the same agent that holds the dangerous tools~\cite{saltzer1975protection,provos2003privilege}.
Our ablation quantifies this: isolation alone provides 323$\times$ improvement; adding structured output formatting closes the gap to zero.

\textbf{Relation to alternative defenses.}
Spotlighting~\cite{hines2024spotlighting} and PromptShield~\cite{jacob2025promptshield} address injection within a single agent via provenance signaling and classifier-based detection respectively.
Progent~\cite{shi2025progent} enforces per-agent tool restrictions via a dynamic policy language.
The Instruction Hierarchy~\cite{wallace2024instruction} trains models to prioritize privileged instructions.
Our approach is complementary: architectural isolation provides a hard constraint that holds independent of model behavior, while IFC-based reasoning~\cite{wu2024ifc} provides a formal lens for the guarantee.
Agent frameworks such as AutoGen~\cite{wu2023autogen} and tool-augmented agents~\cite{yao2023react,schick2023toolformer} make multi-agent decomposition increasingly practical, lowering the deployment cost of our pattern.
Prior work has demonstrated multi-agent LLM pipelines in production planning contexts~\cite{tsao2023multiagent}, validating that role-specialized agent decomposition is a viable engineering pattern beyond research prototypes.

\textbf{Limitations.}
The ablation runs 649 attacks (those that succeeded at baseline), not the full 22,899.
Adaptive attacks designed specifically to survive JSON summarization and multi-agent handoffs are not evaluated; standardized evaluation harnesses such as JailbreakBench~\cite{chao2024jailbreakbench} provide a template for this rigor.
The validator uses simple regex pattern matching; a more sophisticated attacker could craft summaries that evade it while still influencing Agent~2.

\textbf{Residual risk.}
The 63.7\% injection leak rate into Agent~1 summaries (measured in the Pipeline) shows that JSON formatting does not prevent content propagation---it prevents \emph{actionable} content propagation.
If future attack variants can encode executable instructions into structured JSON fields that Agent~2 acts on, the defense degrades.

\section{Conclusion}

We propose a two-mechanism architectural defense against prompt injection in LLM agents: \textbf{agent isolation} (privilege-separated tool access) combined with \textbf{JSON-structured inter-agent communication}.
Tested via OpenClaw on 649 attacks that succeeded against a gpt-5-mini single-agent baseline, our ablation shows:

\begin{itemize}
\item \textbf{JSON formatting alone}: 14.18\% ASR (7.1$\times$ better than baseline)
\item \textbf{Agent isolation alone}: 0.31\% ASR (323$\times$ better than baseline)
\item \textbf{Full pipeline}: 0\% ASR on the evaluated benchmark ($\infty$ improvement)
\end{itemize}

Agent isolation is the dominant mechanism because it imposes a hard structural constraint independent of model behavior.
JSON formatting provides complementary hardening by stripping persuasive framing before it reaches the action agent.
Together, they provide a defense that holds regardless of how convincing any individual injection payload is.

We recommend privilege-separated agent architectures with structured inter-agent communication as a baseline security requirement for production agentic systems handling untrusted input.

\bibliography{references}

\end{document}